\documentstyle[ preprint, 12pt, aps]{revtex}
\newcommand{\be}{\begin{equation}}
\newcommand{\ee}{\end{equation}}
\newcommand{\ba}{\begin{eqnarray}}
\newcommand{\ea}{\end{eqnarray}}
\newcommand{\nn}{\nonumber}
\newcommand{\ov}[1]{\overline{#1}}

\begin{document}
\tightenlines

\begin{center} 

\Large

{\bf 1D lattice dynamics of the diffusion 
limited reaction $ {\bf A+A \rightarrow A +S }$:
Transient behavior}
\vskip1cm
\large
Enrique Abad, \footnote[1]
{ Center for Nonlinear Phenomena and Complex Systems
Universite Libre de Bruxelles,
C.P. 231 Campus Plaine
B-1050 Bruxelles. }
Harry L. Frisch \footnote[2] 
{ Department of Chemistry, State University of New York at Albany,
Albany, New York 12222.}
and Gregoire Nicolis \footnotemark[1] \\[1cm]
\normalsize
{\bf Abstract}\\[.5cm]
\end{center}
We use a boolean cellular automaton model to describe the diffusion limited 
dynamics of the irreversible reaction $ A+A \rightarrow A +S $ on a
1D lattice. We derive a set of equations for the dynamics of the 
empty interval probabilities from which explicit expressions
for the particle concentration and the two point correlation 
 can be obtained. It is shown that the long time dynamics is in agreement   
with the off-lattice solution. The early time behavior, however,
predicts a slower decay of the concentration.\\


\vspace*{2cm}
\noindent
\newpage

\section{Introduction}

In a recent paper \cite{Avr}, ben-Avraham et al. 
treated the 1D dynamics of the 
diffusion limited reaction $A+A\rightarrow A+S$. The analysis
led to an infinite hierarchy of differential equations (HDE) for the
time evolution of the set of quantities $\left\{E_k(t)\right\}$,
where $E_k(t)$ denotes the probability of finding an interval of
$k$ contiguous empty lattice sites 
(c.f. eq. (2.2) and (2.5) in \cite{Avr} ).
Ben-Avraham {\it et al.} were primarily interested in the continuum 
limit for the probability densities $E_k(t)/ (\Delta x)^k $ as 
the lattice spacing $\Delta x$ is allowed to vanish. Here we
rederive the HDE taking as a starting 
point a detailed cellular automaton description using Boolean 
occupation numbers \cite{Aba,Pri}. 
Instead of going to the continuum limit, we solve the HDE for a 
discrete lattice and obtain explicit expressions
for the concentration $c(t)=\left( 1-E_1(t) \right) / \Delta x $
and the two-point nearest neighbour correlation. 
We consider both the case of an initially fully occupied lattice 
and the case in which the initial probabilities $E_k(0)$ are given by 
the random ``geometrical'' distribution $(1-\rho)^k$, where $\rho<1$ 
is the initial probability of a single site being occupied. Some
exact solutions to related problems and a special case can be 
found in \cite{Spou}. 

In section \ref{der}, we derive the HDE starting from the cellular 
automaton dynamical rule. The next section is a brief reminder of
the off-lattice solution for the HDE obtained in \cite{Avr}. In section 
\ref{sodi}, we solve the solution of the HDE for a discrete lattice 
and discuss the transient dynamics of the concentration. As expected, 
the long time behavior is the same as that found in \cite{Avr}. 
However, the early time behavior turns out to be 
significantly different.  
\section{ Derivation of the HDE}
\label{der}
Consider a 1D lattice with $N$ sites. Each lattice site $i$ 
is characterized by a boolean occupation number $n_i=1$ or $0$ 
depending on whether it is occupied by a single particle (A) or 
empty (S).  
At each time step $\delta t$, we choose randomly a site $i$ in the 
lattice via the boolean stochastic parameter $\xi_N^{(i)}$. This 
parameter is equal to one for the chosen site and $0$ for all other
sites. Simultaneously, a second boolean variable is used to select
the left ($\xi_L=1$) or the right neighbour site ($\xi_L=0$) with equal 
probability $1/2$. 
If the site $i$ is occupied, the particle will hop to the 
chosen neighbour site with a rate $k_D$ given by the mean
value of the stochastic boolean parameter $\xi_D$. 
In this case, if the neighbour site is empty, the particle will occupy it 
vacating the site $i$. We express this by including the loss terms
\ba
\label{cont1}
&& -\xi_N^{(i)}(t)\,\xi_L(t)\,\xi_D(t) n_i(t) (1-n_{i-1}(t))
\\
&& -\xi_N^{(i)}(t)\,(1-\xi_L(t))\,\xi_D(t) n_i(t) (1-n_{i+1}(t))
\ea
in the dynamical rule for $n_i(t)$. If the neighbour site is filled, 
the particle will ``react'' with it and instantaneously disappear. 
Again, the occupation number at site $i$ will be decreased 
from $1$ to $0$. Thus, we have the reactive loss terms:
\ba
&& -\xi_N^{(i)}(t)\,\xi_L(t)\,\xi_D(t) n_i(t) n_{i-1}(t) \\
&& -\xi_N^{(i)}(t)\,(1-\xi_L(t))\,\xi_D(t) n_i(t) n_{i+1}(t)
\ea
Finally, as an empty site $i$ can only be occupied by hopping from 
a particle at a neighbour site, one has the two gain terms:
\ba
&&+\xi_N^{(i+1)}(t)\,\xi_L(t)\,\xi_D(t) n_{i+1}(t) (1-n_i(t))
\\
\label{cont6}
&& +\xi_N^{(i-1)}(t)\,(1-\xi_L(t))\,\xi_D(t) n_{i-1}(t) (1-n_i(t))
\ea
Clearly, the dynamics described above corresponds to the particular 
implementation of the reaction $A+A\rightarrow A+S$ given in \cite{Avr}.
Adding up all the contributions (\ref{cont1})-(\ref{cont6}), 
we obtain the following dynamical rule:
\ba
\label{dynrul}
n_i(t+\delta t)&=&n_i(t)-\xi_N^{(i)}(t)\,\xi_D(t) n_i(t) \nn \\
&&+\xi_N^{(i+1)}(t)\,\xi_L(t)\,\xi_D(t) n_{i+1}(t) (1-n_i(t)) \nn \\
&& +\xi_N^{(i-1)}(t)\,(1-\xi_L(t))\,\xi_D(t) n_{i-1}(t)\,(1-n_i(t)),
\quad i=1,\ldots, N. 
\ea 
In the first and last equation (\ref{dynrul}), 
we set $n_0(t)=n_{N+1}(t)=0$.
We can rewrite (\ref{dynrul}) by using the complementary occupation 
numbers $s_i(t)=1-n_i(t)$:
\ba
\label{holdyn}
s_i(t+\delta t)&=&s_i(t)+\xi_N^{(i)}(t)\,\xi_D(t)
-[\xi_N^{(i-1)}(t)+\xi_N^{(i)}(t)]\xi_D(t) s_i(t) \nn \\
&& +\xi_N^{(i-1)}(t)\,(1-\xi_L(t))\,\xi_D(t) s_{i-1}(t)\,s_i(t) \nn \\
&& +\xi_N^{(i+1)}(t)\, \xi_L(t) \,\xi_D(t) s_i(t)\, s_{i+1}(t), 
\qquad i=1,\ldots, N. 
\ea 
This is a more convenient form, since,
as it turns out, the evolution law for a string of $k$ consecutive 
empty sites $\prod_{j=i}^{i+k-1}s_j$ involves only products 
of contiguous occupation numbers
\ba
\label{prodev}
\prod_{j=i}^{i+k-1}s_j(t+\Delta t)&=&\prod_{j=i}^{i+k-1}s_j(t)+
\xi_N^{(i)}(t)\xi_L(t)\xi_D(t)\prod_{j=i+1}^{i+k-1}s_j(t) \nn \\
&& +\xi_N^{(i+k-1)}(t)(1-\xi_L(t)) \xi_D(t)\prod_{j=i}^{i+k-2}s_j(t) \nn \\
&& -[\,\xi_N^{(i-1)}(t)(1-\xi_L(t))+\xi_N^{(i)}(t)\xi_L(t) \nn \\
&&+\xi_N^{(i+k-1)}(t)(1-\xi_L(t))+\xi_N^{(i+k)}(t)\xi_L(t)\,]\,\xi_D(t)
\prod_{j=i}^{i+k-1}s_j(t) \nn \\
&& +\xi_N^{(i-1)}(t)(1-\xi_L(t))\xi_D(t)\prod_{j=i-1}^{i+k-1}s_j(t)
+\xi_N^{(i+k)}(t) \xi_L(t)\xi_D(t) \prod_{j=i}^{i+k}s_j(t)
\ea
Let us take the time step $\delta t$ equal to $1/N$, implying that each 
site has been visited once on average after one time unit. 
If we now average (\ref{prodev}) over an ensemble of realizations for
a given initial configuration, we obtain in the thermodynamic limit 
$N\rightarrow\infty$:
\be
\frac{dE_k^i}{dt}=\frac{k_D}{2}\,\left(E_{k-1}^{i+1}+
E_{k-1}^i-4E_k^i+E_{k+1}^{i-1}+E_{k+1}^i\right), 
\ee
where $E_k^i(t)=\overline{\prod_{j=i}^{i+k-1}s_j(t)}$. For a single 
site ($k=1$) the corresponding equation 
\be
\frac{dE_1^i}{dt}=k_D\,\left(1-2E_1^i+\frac{1}{2}E_2^{i-1}
+\frac{1}{2}E_2^i\right)
\ee
is obtained by averaging the dynamical rule (\ref{holdyn}).
If we perform the average over realizations \underline{and} translationally
invariant initial conditions, we get the following hierarchy of 
differential difference equations for the evolution of the averaged products 
$E_k(t)=\langle \prod_{j=i}^{i+k-1}s_j(t) \rangle$:
\be
\label{hier}
\frac{dE_k}{dt}= k_D \left(E_{k+1}-2E_k+E_{k-1}\right), \qquad k=1,2,\ldots 
\ee
with the boundary condition $E_0(t)=1$. As described in \cite{Avr}, the
rhs of (\ref{hier}) represents the net flux due to particle diffusion 
into and out of an empty site interval, whereas the effect of reaction 
enters through the boundary condition.  
\section{Off-lattice solution}
\label{offl}

Following ref. \cite{Avr}, we set the hopping rate $k_D/2$ to 
either of both sides equal to $D/ (\Delta x)^2$, where 
$\Delta x$ is the lattice spacing. On long length 
and time scales this yields normal diffusion with a diffusion 
coefficient $D$. With this definition, eqs. (\ref{hier}) are
identical to those derived in \cite{Avr}. If we now let 
$\Delta x\rightarrow 0$, eqs. (\ref{hier}) become
\be
\label{ediff}
\frac{\partial E(x,t)}{\partial t}= 2D 
\,\frac{\partial^2 E(x,t)}{\partial x^2}.
\ee
with the boundary conditions $E(0,t)=1$ and $E(\infty,t)=0$. 
In this limit, the concentration (number of particles
per unit length) is expressed as
\be
c(t)=-\left.\frac{\partial E(x,t)}{\partial x}\right|_{x=0}
\ee  
Thus, one can determine the time dependence of the concentration 
by solving (\ref{ediff}) with the boundary conditions given above 
(see \cite{Avr} for details). For the special case of an initially
random particle distribution with a concentration $c_0$, one
has
\be
\label{baerl}
\frac{c(t)}{c_0}=1-
\left(\frac{8c_0^2Dt}{\pi}\right)^{\frac{1}{2}}+o(c_0^2Dt)
\quad\mbox{as }t\rightarrow 0
\ee 
and
\be
\label{lta}
c(t)\rightarrow \frac{1}{\left(2\pi Dt\right)^{\frac{1}{2}}}\quad\mbox{as }
t\rightarrow \infty
\ee
\section{Solution for a discrete lattice}
\label{sodi}
We now proceed to solve the HDE (\ref{hier}) for the case in which one
has an initially full lattice, i.e. $E_k(0)=0$ for all $k\ge 1$. To begin 
with, we absorb the rate constant $k_D$ into the time scale by introducing  
the dimensionless time variable $\tau=k_D \, t$. The hierarchy then reads:
\be
\label{resc}
\frac{dE_k}{d\tau}=E_{k+1}-2E_k+E_{k-1}, \qquad k=1,2,\ldots
\ee
Next we apply the Laplace transform to both sides of (\ref{resc}) and obtain
the homogeneous difference equation
\be
\label{diffeq}
\ov{E}_{k+1}-(2+s)\ov{E}_k+\ov{E}_{k-1}=0, \qquad k\ge 1,
\ee
where $\ov{E}_k(s)=L_{\tau \rightarrow s}\left\{E_k(\tau)\right\}=
\displaystyle{\int_0^{\infty}\exp(-s\tau)\,E_k(\tau)\,d\tau}$.
The boundary condition is given by $\ov{E}_0(s)=1/s$.
This second-order difference equation is solved with the ansatz 
$E_k(s)=\lambda^k(s)$. This leads to the quadratic equation
\be
\lambda^2-(2+s)\lambda+1=0
\ee
which has the two solutions 
\be
\lambda_{\pm}=\frac{s+2\pm\sqrt{s^2+4s}}{2}
\ee    
The general solution of (\ref{diffeq}) is obtained as a linear 
superposition of $\lambda_+^k$ and $\lambda_-^k$:
\be
\label{gensol}
\ov{E}_k(s)=A(s)\,\lambda_-^k(s)+B(s)\,\lambda_+^k(s)
\ee
For $k\rightarrow \infty$, the physically acceptable solution of 
(\ref{diffeq}) must satisfy 
the implicit boundary condition $E_{\infty}(s)=0$. To avoid
the divergence of the second term in (\ref{gensol}), we must therefore 
set $B(s)=0$. 
Using the other boundary condition at $k=0$, we find 
$A(s)=1/s$. Thus, we
have:
\be
\label{lapset}
\ov{E}_k(s)=\frac{1}{s}\left(\frac{s+2-\sqrt{s^2+4s}}{2}\right)^k
\ee  
Clearly, the most interesting quantity is 
$\ov{E}_1(s)$, whose inverse Laplace
transform $E_1(\tau)$ is the probability that a randomly chosen
 site be empty. 
By virtue of a theorem \cite{Car},
the inverse transform 
$L_{s\rightarrow\tau}^{-1}\{\ov{E}_k(s)\}$ is given by
the integral
\be
\int_0^{\tau} v_k(\tau')\,d\tau', \qquad k=1,2,\ldots.
\ee 
where
\be
v_k(\tau)=L_{s\rightarrow\tau}^{-1}
\left\{\left(\frac{s+2-\sqrt{s^2+4s}}{2}\right)^k\right\}=
k\,\,\frac{\exp(-2\tau)\,I_k(2\tau)}{2\tau}
\ee
(see e.g. \cite[page 379]{Cra}), where 
\be
\label{bess}
I_n(x)=\sum_{r=0}^{\infty}\frac{(x/2)^{2r+n}}{r!\, \Gamma(r+n+2)}
\ee  
are the modified Bessel functions. In particular, one has 
\be
\label{hocon}
E_1(\tau)=\int_0^{\tau} \frac{\exp{(-2\tau')} I_1(2\tau')}{\tau'}\, d\tau'.
\ee
\subsection{Asymptotics for early times}
For sufficiently short times ($\tau \ll 1$), we can use (\ref{bess}) and 
the series 
expansion of the exponential function $\exp{(-2\tau')}$ to expand the 
integrand in (\ref{hocon}) 
in powers of $\tau'$. Neglecting terms of order $o(\tau'^3)$,  
performing the 
integration and undoing the time scaling, we obtain:
\be
\label{exp}
E_1 (t)=k_D\,t-k_D^2\,t^2+\frac{5}{6}\,k_D^3\,t^3+o(t^4)
\ee
The particle concentration is 
\be
c(t)=\frac{P_1(t)}{\Delta x}=\frac{1-E_1(t)}{\Delta x}=
c_0 \left[ 1-2c_0^2Dt+4c_0^4 D^2 t^2+o(c_o^6 D^3 t^3)\right],
\ee
where $c_0=1/\Delta x$. This is in clear disagreement with the 
decay law (\ref{baerl}). 

For sufficiently short times, we can 
neglect the effect of large clusters of
vacant sites, since the chain is initially full. This is done by 
setting $E_k=0$ 
for $k$ larger than a certain cutoff size $k_c$ in the truncation 
hierarchy (\ref{resc}). If we set $k_c=1$, 
we obtain the differential equation
\be
\label{trunc}
\frac{dE_1}{d\tau}=1-2E_1(\tau).
\ee
The solution reads
\be
E_1(\tau)=\frac{1}{2}\left(1-\exp{(-2\tau)}\right)=\tau-\tau^2+
\frac{2}{3}\tau^3+o(\tau^4) 
\ee 
Setting $\tau=k_D\,t$, we see that the early times expansion (\ref{exp}) is
reproduced correctly up to the quadratic term. The discrepancy between 
(\ref{exp}) and the off-lattice solution arises due to the 
finite propagation velocity of a local perturbation in concentration, as
opposed to the infinite propagation velocity characteristic of 
diffusion. Thus, we expect a slower decay of the concentration $c(t)$ 
on the lattice. 
\subsection{Long time asymptotics} 
For large times ($\tau \gg 1$), we write $E_1(\tau)$ as follows:
\be
\label{ltexp}
E_1(\tau)=A-\int_{\tau}^{\infty} \frac{\exp{(-2\tau')} 
I_1(2\tau')}{\tau'}\, d\tau'.
\ee
where $A$ is the definite integral
\be
\int_0^{\infty} \frac{\exp{(-2\tau')} I_1(2\tau')}{\tau'}\, d\tau',
\ee
which is equal to 1 \cite[page 236]{Rish}. 
The integrand in the second term of (\ref{ltexp}) 
can be expanded using the asymptotic form
\be
I_1(x)=\frac{\exp{(x)}}{\sqrt{2\pi x}}\left(1-\frac{3}{8 x}
+o\left(\frac{1}{x^2}\right)\right).
\ee
for large $x$ ( see e.g. \cite[page 489]{Car} ). Thus, we get
\ba
E_1(\tau)&=&1-\frac{1}{2 \sqrt{\pi}}\int_{\tau}^{\infty}
\tau'^{-\frac{3}{2}}\,d\tau'
+\frac{3}{32 \sqrt{\pi}}\int_{\tau}^{\infty}
\tau'^{-\frac{5}{2}}\,d\tau'+o(\tau^{-\frac{5}{2}}) 
\nn \\
&=&1-\frac{1}{\sqrt{\pi\tau}}+
\frac{1}{16 \sqrt{\pi\tau^3}}+o(\tau^{-\frac{5}{2}}).
\ea
This is in agreement with the long time asymptotics of 
the continuum limit solution (\ref{lta}). 
\subsection{Two-point correlation }
The explicit expression for the two interval probability
\be
E_2(\tau)=2\,\int_0^{\tau}
\frac{\exp{(-2\tau')} I_2(2\tau')}{\tau'}\, d\tau'.
\ee  
can be used as a starting point to compute the asymptotics
of the two-point nearest neighbour correlation 
$c^{(2)}(t)=P_2/ (\Delta x)^2$, where $P_2=1-2E_1+E_2$ is the 
probability of finding two contiguous particles in the lattice. 
For early times one gets 
$c^{(2)}(t)=c_0^2\,\left[1-c_0^2Dt+o(c_0^4D^2t^2)\right]$,
whereas for long times  
$c^{(2)}(t)\approx 1/c_0\sqrt{32\pi}(Dt)^{3/2}$.  
\subsection{Solution for an arbitrary initial concentration} 
In this case, we average over all possible initial configurations of 
the lattice with a given concentration $c_0=\rho/\Delta x$ ($\rho<1$). 
The initial conditions for the empty interval probabilities now read
\be
E_k(0)=(1-\rho)^k, \qquad k=1,2,\ldots
\ee
The boundary conditions are the same as in the case of an initially
 full lattice. 
If we now apply
the Laplace transform to (\ref{resc}), we obtain 
\be
\label{diffeq2}
\ov{E}_{k+1}-(2+s)\ov{E}_k+\ov{E}_{k-1}=-(1-\rho)^k, \qquad k\ge 1,
\ee
which differs from (\ref{diffeq}) by the inhomogeneity on the rhs. 
Like in the theory
of ordinary differential equations, the general solution of (\ref{diffeq2})
can be written as the sum of the
general solution (\ref{gensol}) for (\ref{diffeq}) 
and a particular solution which we seek in the form 
\be
\ov{E}_k^{par}(s)=(1-\rho)^k\,C(s).
\ee
Inserting this ansatz in (\ref{diffeq2}), we find 
\be
\label{const}
C(s)=\frac{(1-\rho)}{(1-\rho)s-\rho^2}.
\ee
The boundary condition for $k\rightarrow \infty$ again imposes $B(s)=0$. 
From the 
boundary condition for $\ov{E}_0(s)$ we obtain
\be
\label{const2}
A(s)=\frac{1}{s}-\frac{1-\rho}{(1-\rho)s-\rho^2}.
\ee
Putting (\ref{const}) and (\ref{const2}) into the equation 
\be
\ov{E}_k(s)=A(s)\lambda_-^n(s)+\ov{E}_k^{par}
\ee
we find
\be
\label{lapE}
\ov{E}_k(s)=\left(\frac{1}{s}-\frac{1-\rho}{(1-\rho)s-\rho^2}\right)\left[
\frac{s+2-\sqrt{s^2+4s}}{2}\right]^k+\frac{(1-\rho)^{k+1}}{(1-\rho)s-\rho^2}, 
\quad k=0,1,\ldots
\ee
We can now use the convolution theorem for the Laplace 
transform \cite{Car} to invert (\ref{lapE}). This yields
\ba
E_k(\tau)&=&k\,
\int_0^{\tau} \frac{\exp{(-2\tau')} I_k(2\tau')}{\tau'}\, d\tau'
\nn \\
&&+\left[\left(1-\rho\right)^k-k\,\int_0^{\tau} 
\frac{\exp{\left(-\left[2+\frac{\rho^2}{1-\rho}\right]\,\tau'
\right)} 
I_k(2\tau')}{\tau'}\, d\tau'\right]\,\exp{\left(\frac{\rho^2\,\tau}
{1-\rho}\right)}.
\ea
 However, we can study the asymptotics directly from 
(\ref{lapE}) by making use of the Tauberian theorems \cite{Doe}, 
which allow us to determine the behaviour of 
$E_1(\tau)$ for $\tau\rightarrow 0$ and 
$\tau\rightarrow\infty$ by inverting respectively the series expansion of 
$\ov{E}_1(s)$ around $s=\infty$ and $s=0$. For small $s$ one has
\be
\ov{E}_1=\frac{1}{s}-\frac{1}{\sqrt{s}}+
\frac{1}{2}+\frac{1-\rho}{\rho}+o(s^{\frac{1}{2}})
\ee
from which we get
\be
E_1\approx 1-\frac{1}{\sqrt{\pi\tau}} 
\qquad \mbox{for}\quad \tau\rightarrow \infty.
\ee
Again, this is in agreement with $(\ref{lta})$. The long time asymptotics 
does not depend on the initial concentration $c_0$, suggesting
that a universal behavior also takes place on a finite lattice. 
In the opposite limit we have
\be
\label{lap1}
\ov{E}_1=\frac{1-\rho}{s}+\frac{\rho^2}{s^2}
-\frac{\rho^2(1+\rho)}{s^3}+o(s^4)
\ee
Inverting term by term the rhs of (\ref{lap1}), we get
\be
\label{arber}
E_1=1-\rho+\rho^2\tau-\frac{\rho^2(1+\rho)}{2}\tau^2+o(\tau^3)
\ee
leading to 
\be
c(t)=c_0 \left[ 1-\frac{2}{\rho}c_0^2Dt+
2\frac{\rho^2(\rho+1)}{\rho^5}c_0^4 D^2 t^2+o(c_0^6 D^4 t^4)\right]
\ee
Thus, the discrepancy with (\ref{baerl}) appears to be robust. 
We can again compare the exact early time expansion (\ref{arber}) 
with the solution of the differential 
equation (\ref{trunc}) with the initial condition $E_1(0)=1-\rho$, 
which is given by 
\be
\frac{1}{2}+\left(\frac{1}{2}-\rho\right)\exp{(-2\tau)}=
1-\rho+(2\rho-1)\tau+(1-2\rho)\tau^2+o(\tau^3).
\ee
As expected, the approximation based on the neglect of empty intervals 
becomes worse as $\rho$ decreases. 
\section{Acknowledgments}
This work was supported, in part, by NSF grant DMR 9628224, the
Training and Mobility of Researchers program of the European Commission
and by the Interuniversity Attractions Poles program of the 
Belgian Federal Government. E. Abad acknowledges helpful discussions
with F. Vikas and F. Baras.


\begin{thebibliography}{99}
\bibitem{Avr} D. ben Avraham, M. Burschka and C. Doering,
 J. Stat. Phys. {\bf 60} 695 (1990).
\bibitem{Aba} E. Abad, G. Nicolis and P. Grosfils, {\it 
Nonlinear reactive systems viewed as stochastic dynamical 
systems}, unpublished.
\bibitem{Pri} V. Privman, Phys. Rev. E {\bf 50}, 50 (1994).
\bibitem{Spou} See J.L. Spouge, Phys. Rev. Lett. {\bf 60} 871 (1988)
when initial ``concentration'' $\rho$ is $1/2$ ; for the case 
with reverse reaction see 
 J.C. Lin, C.R. Doering and D. ben-Avraham, Chem. Phys. 
{\bf 146}, 355 (1990). There exist some exact solutions of 
the equivalent diffusion-annihilation process 
$A+A \rightarrow S+S$, see e.g.
R.J. Glauber, J. Math. Phys. {\bf 4}, 294 (1963);
T. Liggett, {\it Interacting Particle Systems}
(Springer Verlag, New York, 1985); 
S. Redner and K. Kang, Phys. Rev. A {\bf 32}, 435 (1985);
A.A. Lushnikov, Phys. Lett. A {\bf 120}, 135 (1987);
V. Kuzokov and E. Kotomin, Rep. Prog. {\bf 87}, 1941
(1988);  
M. Bramson and J. Lebowitz, Phys. Rev. Lett. {\bf 61}, 2397
(1988);
D. Balding, P. Clifford and N.J.B. Green, Phys. Lett. A {\bf
126}, 481 (1988);
F. Family and J.G. Amar, J. Stat. Phys. {\bf 65}, 1235 (1991);
M.D. Grynberg and Robin B. Stinchcombe, Phys. Rev. Lett. 
{\bf 76}, 851 (1996)
and references therein.       
\bibitem{Cra} J. Crank, {\it The mathematics of diffusion},
2nd edition reprinted, Clarendon, Oxford (1976).
\bibitem{Rish} I.M. Ryshik and I.S. Gradstein, 
{\it Tables of series, products
and integrals}, 2nd edition, VEB Deutscher Verlag der Wissenschaften, 
Berlin (1963).
\bibitem{Car} H.S. Carslaw and J.C. Jaeger, {\it Conduction of heat in 
solids}, 2nd edition, Clarendon, Oxford (1959).
\bibitem{Doe} G. Doetsch, {\it Theorie und Anwendung der Laplace
Transformation}, Dover Pub., New York (1943).
\end{thebibliography}
\end{document}